\documentclass[sigconf]{acmart}
\usepackage{bbold}
\AtBeginDocument{%
  }

\copyrightyear{2025}
\acmYear{2025}
\setcopyright{cc}
\setcctype{by}
\acmConference[UbiComp Companion '25]{Companion of the 2025 ACM International Joint Conference on Pervasive and Ubiquitous Computing}{October 12--16, 2025}{Espoo, Finland}
\acmBooktitle{Companion of the 2025 ACM International Joint Conference on Pervasive and Ubiquitous Computing (UbiComp Companion '25), October 12--16, 2025, Espoo, Finland}\acmDOI{10.1145/3714394.3756220}
\acmISBN{979-8-4007-1477-1/2025/10}

\begin{document}

\title{Passive RF Sensing for People Counting with Dense IoT Networks}


\author{Federica Fieramosca}
\affiliation{%
 \institution{Politecnico di Milano}
 \city{Milan}
 \state{}
 \country{Italy}}
 \email{federica.fieramosca@polimi.it}

\author{Vittorio Rampa}
\affiliation{%
  \institution{Consiglio Nazionale delle Ricerche}
  \city{Milan}
  \state{}
  \country{Italy}}
  \email{vittorio.rampa@ieiit.cnr.it}

\author{Michele D'Amico}
\affiliation{%
  \institution{Politecnico di Milano}
 \city{Milan}
 \state{}
 \country{Italy}}
\email{michele.damico@polimi.it}

\renewcommand{\shortauthors}{Federica Fieramosca, Vittorio Rampa, \& Michele D’Amico}
\begin{abstract}

Radio Frequency (RF) convergence, the joint use of radio signals for both communication and sensing, is gaining increasing attention in research and industry, mostly for its potential in Internet of Things (IoT) applications. By using RF signals of opportunity, such as those generate by the wireless networks, RF sensing exploits the effects of human bodies on the electromagnetic waves to detect both people presence and their movements without requiring wearable devices.

In this paper, we address the problem of estimating the number of people in a monitored area covered by several RF nodes of a wireless network: to this aim, we propose a composite diffraction-based model, namely C-MAM, to capture the cumulative RF attenuation induced by multiple human bodies. Building on this model, we derive the analytical resolvability bounds that quantify the maximum number of distinguishable individuals based on the layout geometry and other physical characteristics of the wireless network. These bounds allow for the pre-deployment assessment of RF sensing performance under different scenarios. 

Theoretical findings are validated through simulations and experimental measurements, confirming the model’s predictive accuracy and practical relevance.

\end{abstract}





\maketitle

\section{Introduction}

Radio Frequency (RF) sensing refers to a set of opportunistic techniques designed to detect, locate, and track people within areas covered by ambient RF signals~\cite{wilson-2010}. In the context of the \emph{Integrated Sensing and Communication} (ISAC) paradigm~\cite{savazzi-1}, RF sensing systems turn each wireless node into a \emph{virtual sensor}, enabling simultaneous communication and sensing. Typically operating in the unlicensed $2.4\div6$~GHz ISM bands, these systems exploit cm-scale wavelengths to capture body-induced alterations in multipath propagation.

\begin{figure}[h]
  \centering
  \includegraphics[width=\linewidth]{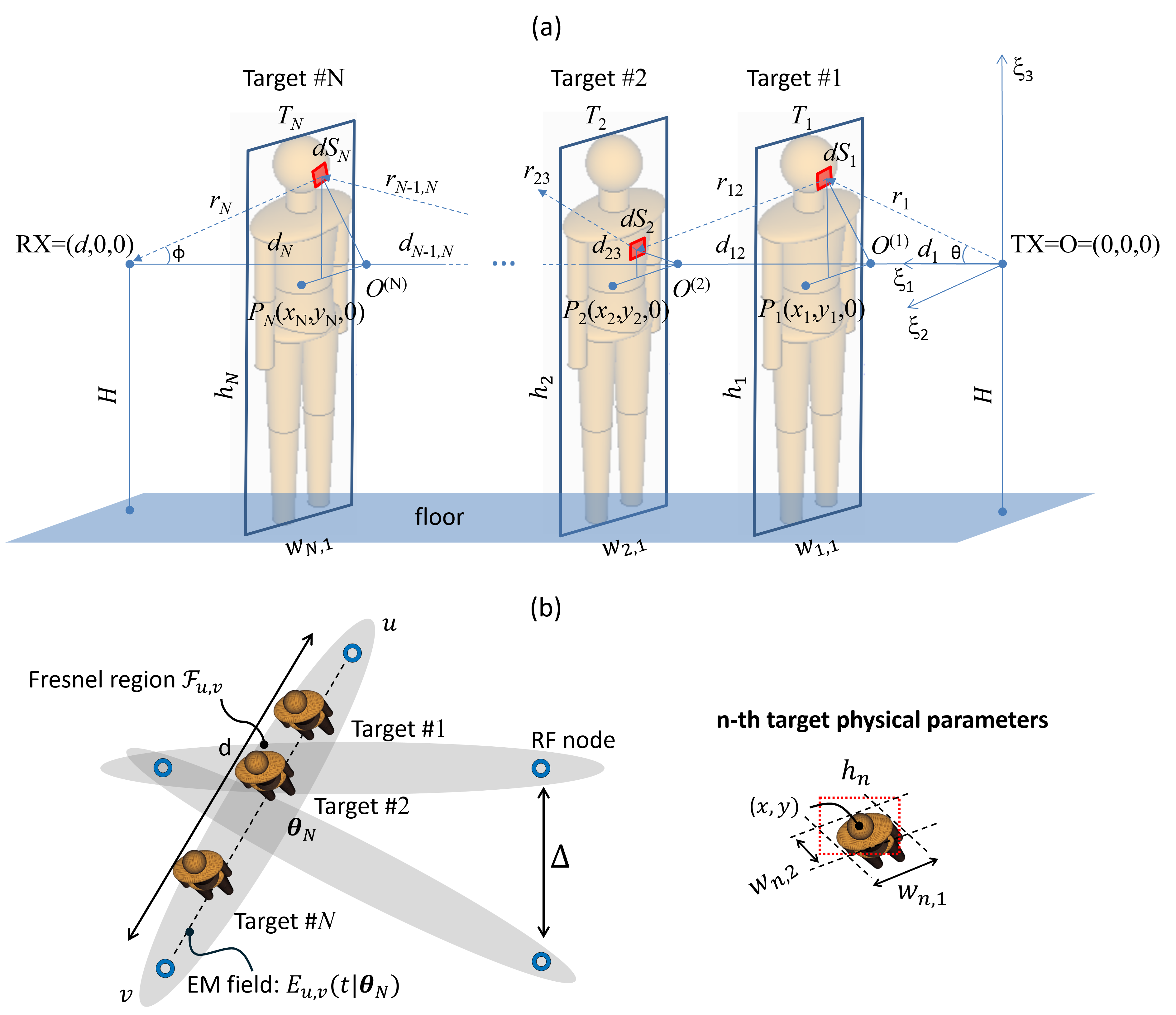}
  \caption{RF sensing scenarios: a) Single-link, multi-target setup with a horizontal link of length $d$ at height $H$, and $N$ targets with footprints at positions $\mathbf{P}_n = (x_n, y_n)$; b) general case with multiple links. Each target is defined by physical features $\theta_n = \left\{ \mathbf{P}_n, \varphi_n, h_n, w_{n,1}, w_{n,2} \right\}$ and randomly placed in the monitored area.}\label{new_intro}
\end{figure}

Sensing techniques such as Bayesian filtering, tomographic imaging, and RF holography~\cite{kat,fall,holog} require a detailed understanding of how human bodies (i.e., the targets) block or scatter EM waves, which limits the applicability of standard supervised Machine Learning~(ML) methods. Furthermore, large-scale RF sensing involves solving ill-posed inverse problems on high-dimensional graph-like data structures~\cite{geom}.

Recent work addresses these challenges by developing physics-based models of body-wave interaction, including ray tracing~\cite{phyindoor,rayback}, point scattering~\cite{scatt}, and diffraction-based approaches like GTD~\cite{james-2007}, UTD~\cite{plouhinec-2023,qi-2017}, and scalar diffraction theory~\cite{rampa-2017,rampa-2022a,j-fieramosca-2023}. Moreover, Generative Neural Networks (GNNs) have been used to accelerate simulations and to tackle inverse imaging tasks~\cite{generative_mag,generation,eucap24}.

\subsection{RF Sensing Scenario: Infrastructural Snapshot}
\label{subsec:RF_sensing_scenario}

We consider a dense deployment of RF nodes that collect measurements to form an \emph{infrastructural snapshot}~\cite{moral} of the environment. The goal is to estimate a snapshot at time $t$ of the number $N(t)$ of human targets within a monitored area $\mathcal{X}$ and characterize their motion and physical traits. Each target has distinct parameters $\boldsymbol{\theta}_N = [\theta_1, \ldots, \theta_N]$, where $\theta_n = \{ \mathbf{P}_n, \varphi_n, h_n, w_{n,1}, w_{n,2} \}$ includes the footprint position $\mathbf{P}_n=(x_n,y_n)$, orientation $\varphi_n$, height $h_n$, and 3D lateral body size $w_n$ such that $w_{n,2} \le w_n \le w_{n,1}$~\cite{generation}.

\subsection{Contributions}

The paper introduces a physically-based EM model, called the \emph{Composite Multi-body Additive Model} (C-MAM), to describe the cumulative attenuation caused by multiple human bodies. Compared to prior approaches~\cite{eibert,smith,pravadelli,depatla,yang}, C-MAM includes occlusion and interference effects within Fresnel's zones and it is derived from classical diffraction theory.

Building on C-MAM, we derive \emph{resolvable target bounds} theoretical limits on the number of distinguishable targets in dense RF networks. These bounds are based on the link geometry, Fresnel's zones overlap, signal frequency, and subject size. They are computed using stochastic geometry and provide guidance for system design and deployment. For related work on single-target pre-deployment strategies, see also~\cite{kianoush-2016}.

The remainder of the paper is structured as follows: Sect.~\ref{sec:syst} presents the system model and network representation. Sect.~\ref{sec:em_model} introduces the C-MAM model, while Sect.~\ref{sec:limits} derives resolvability bounds. Simulation and experimental validations are reported in Sect.~\ref{sec:Numerical-results-and}. Conclusions and future directions are discussed in Sect.~\ref{sec:Conclusions}.

\section{System model\label{sec:syst}}

The monitored environment $\mathcal{X}$ is covered by a set $\mathcal{V} $ of RF nodes (or antennas), placed to ensure full spatial coverage. The set of active communication links is denoted by $\mathcal{L} $, with cardinality $L \leq |\mathcal{V}|(|\mathcal{V}| - 1)$. Together, the nodes and the links define a graph structure $\mathcal{G} = (\mathcal{V}, \mathcal{L})$, where $\mathcal{V}$ represents the set of vertices and $\mathcal{L}$ the set of edges. Each edge corresponds to a physical radio link $\ell=(u, v) \in \mathcal{L}$ between two distinct nodes $u$ and $v$ (i.e., $u \ne v$). In the following sections, $\ell$ and $(u, v)$ will be indifferently used to indicate the link between the nodes $u$ and $v$.

Each node $u \in \mathcal{V}$ is located at position $P_u = (x_u, y_u, z_u)$ and operates either as a transmitter (TX) or receiver (RX), depending on the network’s MAC configuration. The set of neighbors for node $u$ is denoted by $ \mathcal{N}(u) :=\{ v \in \mathcal{V} : (u, v) \in \mathcal{L} \}. $

Communication is governed by a time-division access scheme that allows each node $u$ to sequentially receive RF beacon signals from its neighbors. At each time instant $t$, the node collects a set of EM field measurements:
\begin{equation}
\mathbf{E}_{u,t}(\boldsymbol{\theta}_N) := \left\{ E_{u,v}(t | \boldsymbol{\theta}_N) \right\}_{v \in \mathcal{N}(u)} \in \mathbb{R}^{|\mathcal{N}(u)|}, \label{eq:features}
\end{equation}
where each scalar $E_{u,v}(t| \boldsymbol{\theta}_N)$ represents the received EM signal strength from node $v$ to $u$, possibly affected by the presence of $N$ human targets, whose physical parameters $\boldsymbol{\theta}_N$ are defined in Sect.~\ref{subsec:RF_sensing_scenario}. The vector $\mathbf{E}_{u,t}$ captures the local EM signature observed by node $u$ at time $t$. By aggregating all node-level measurements, a full-system sensing matrix is formed:
\begin{equation}
\mathbf{E}_{t}(\boldsymbol{\theta}_N) = \left\{ \mathbf{E}_{u,t}(\boldsymbol{\theta}_N) \right\}_{u \in \mathcal{V}}, \label{eq: node features}
\end{equation}
which encodes the global EM state of the network at time $t$ under the influence of the $N$ targets. To simplify the geometry, and without loss of generality, we assume that all nodes lie on a horizontal plane parallel to the floor at a constant height $H$, as shown in Fig.~\ref{new_intro}. Therefore, we set $z_u = 0 \hspace{0.1cm} \text{for all } u \in \mathcal{V}$ and drop $z_u$ in the following sections.

Note that, in the following sections, the attenuation values will be evaluated w.r.t. the reference scenario with no people inside the monitored area (i.e., the free-space case). For this reason, the terms \textit{extra attenuation} will be implicitly assumed, unless noted, when referring to the plain term attenuation. The explicit modeling of field perturbations $E_{u,v}(t|\boldsymbol{\theta}_N)$ in terms of body-induced attenuations $A_{u,v}$ will be addressed in the following section.

\section{EM Modeling of Multi-Body Attenuation\label{sec:em_model}}

In this section, we summarize the electromagnetic models that describe the human-body-induced attenuation in RF sensing systems. Starting from the single-target case, the models are extended to multi-subject settings, culminating in the proposed Composite Multi-body Additive Model (C-MAM)~\cite{rampa-2017,rampa-2022a,geom}.

Each generic target is identified by $T_n$ and all targets by the set $\mathcal{T}:=\{n=1,\dots,N: T_n\}$. In the multiple targets scenario of Fig.~\ref{new_intro}.a, $N$ targets $T_n\in\mathcal{T}$ are deployed in the link area. On the contrary, in the single-target scenario, only one target is present: in this case, it is $n=1=N$ and the unique target is identified by $T_{1}$ or, equivalently, by $T_n$ but assuming $n=1$ as well.

The Single-target Model is based on the scalar diffraction theory~\cite{rampa-2017} and evaluates how a single human body, modeled as a 2D EM absorbing sheet $T_n$ ($n=1$), perturbs the EM field between two RF nodes $u$ and $v$. The received field $E_{u,v}(t | \theta_n)$ is expressed as:
\begin{equation}
E_{u,v}(t|\theta_n) = E_{0} \left\{1- j\,\frac{d}{\lambda} \int_{T_n} \frac{1}{r_1 r_2} \exp\left[-j\frac{2\pi}{\lambda}(r_1 + r_2 - d)\right]\right\} dT,
\end{equation}
where $E_{0}$ is the free-space electric field received by the node $u$ when no target is inside the area of the link $(u,v)$, $d$ is the length of the link $(u,v)$, $\lambda$ the wavelength, and $r_1$, $r_2$ the distances from the elementary surface $dT$ to the TX and RX, respectively. The corresponding attenuation $A_{u,v}$ (in dB) is set to $A_{u,v}=A_{T_n} $ where:
\begin{equation}
A_{T_n} = -10\log_{10} \left| \frac{E_{u,v}(t|\theta_n)}{E_0} \right|^2.
\label{eq:AT_dB}
\end{equation}

The Multi-body Additive Model (MAM) generalizes this to multiple targets ($n=1,\ldots,N$) by summing the attenuation contributions of all targets~\cite{rampa-2022a}:
\begin{equation}
A_{u,v} = \sum_{n=1,\ldots,N} A_{T_n},
\end{equation}
where $A_{T_n}$ is the attenuation (\ref{eq:AT_dB}) due to each target $T_n$ placed in the link area. For each node $u$, the attenuation values across neighboring links form the feature vector:
\begin{equation}
\mathbf{A}_{u,t} := \left\{ A_{u,v} \right\}_{v \in \mathcal{N}(u)},
\end{equation}
and the network-wide matrix is:
\begin{equation}
\mathbf{A}_t = \left\{ \mathbf{A}_{u,t} \right\}_{u \in \mathcal{V}}.
\end{equation}

In dense scenarios, additive models like MAM may lead to physically implausible attenuation estimates when multiple targets occlude the same region. To overcome the limitations of MAM, the Composite Multi-body Additive Model (C-MAM) introduces two key assumptions~\cite{geom}:
\begin{itemize}
\item A target contributes attenuation only if it lies within the first Fresnel's zone $\mathcal{F}_{u,v}$ of the $\ell$-th link.
\item If one target fully obstructs $\mathcal{F}_{u,v}$, further targets do not add attenuation.
\end{itemize}
The modified per-target attenuation becomes:
\begin{equation}
A_{T_n} = 
\begin{cases}
-10 \log_{10} \left| \frac{E_{u,v}(t,\theta_n)}{E_0} \right|^2, & \text{if } T_n \in \mathcal{F}_{u,v}, \\
0, & \text{otherwise},
\end{cases}
\end{equation}
and the total attenuation is:
\begin{equation}
A_{u,v} =
\begin{cases}
\max \left\{ A_{T_n} \right\}, & \text{if } \exists\, T_n \in \mathcal{F}_{u,v}, \\
0, & \text{otherwise}.
\end{cases}
\end{equation}

C-MAM prevents non-physical overestimation of the attenuation by avoiding additive contributions in shadowed regions, offering a more realistic representation of multi-target RF propagation effects in dense environments~\cite{rampa-2017}.

\begin{figure}[h]
  \centering
  \includegraphics[width=\linewidth]{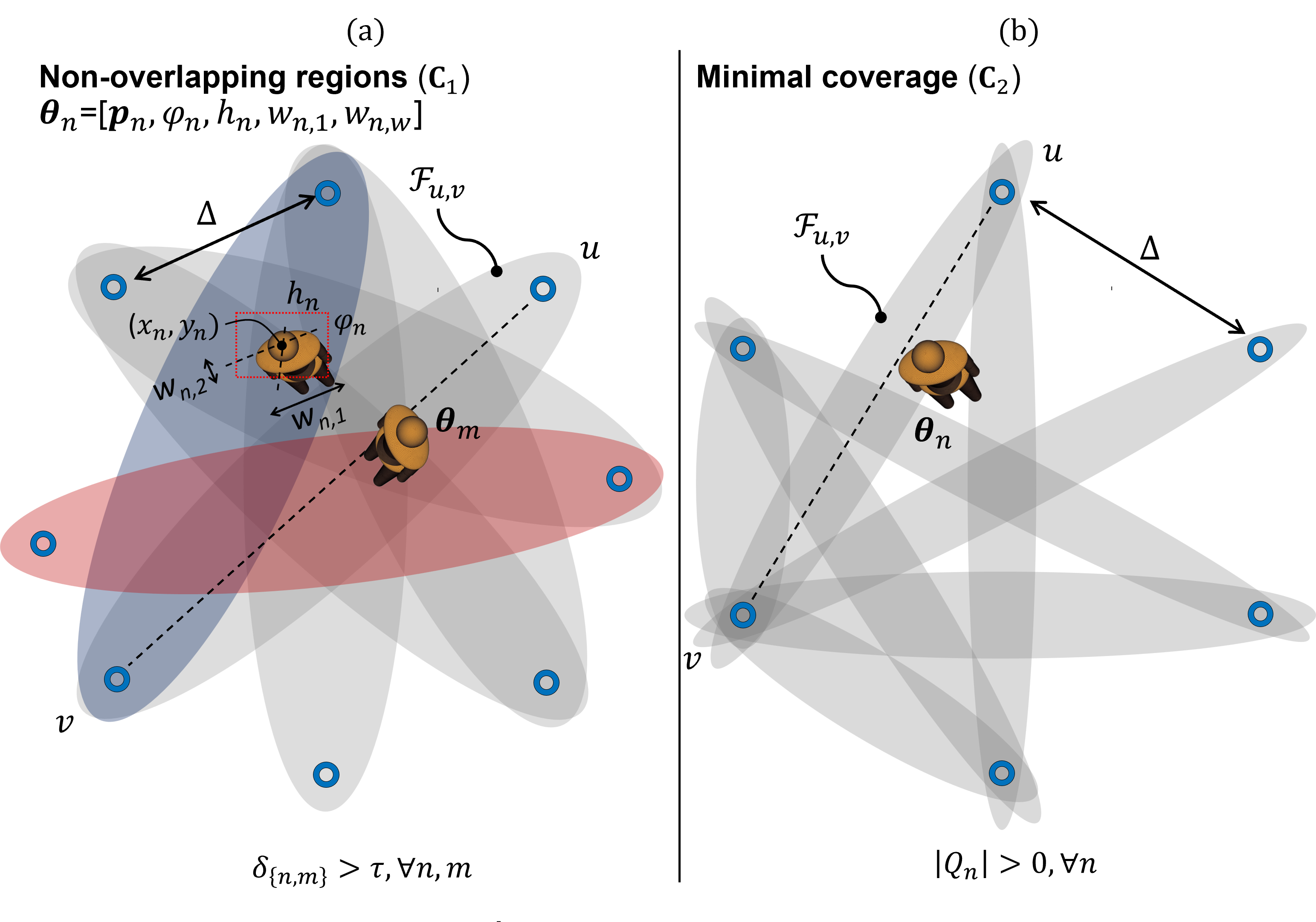}
  \caption{ \label{c1c2} a) Condition $\mathbf{C1}$: a target $T_n$ is resolvable if the Fresnel's zones $Q_n$ and $Q_m$ covering targets $n$ and $m$ are sufficiently distinct, i.e., $\delta_{n,m} > \tau$ (Jaccard distance). Blue and red regions illustrate non-overlapping zones. b) Condition $\mathbf{C2}$: target $T_n$ is unresolved as it lies outside all Fresnel's zone ($|Q_n| = 0$), i.e. no link fully covers it.}
\end{figure}

\section{Limits on resolvable subjects in dense graphs\label{sec:limits}}

This section establishes a theoretical framework to compute the maximum number of human targets that can be reliably resolved using a dense RF sensing network. The analysis is grounded in diffraction theory and leverages the spatial structure of Fresnel's zones to define resolvability conditions. The proposed bound serves as a tool to assess the impact of network density, link layout, and signal frequency on the target resolution capacity of the system.

\begin{figure*}
\centering
\includegraphics[width=\textwidth,scale=0.5]{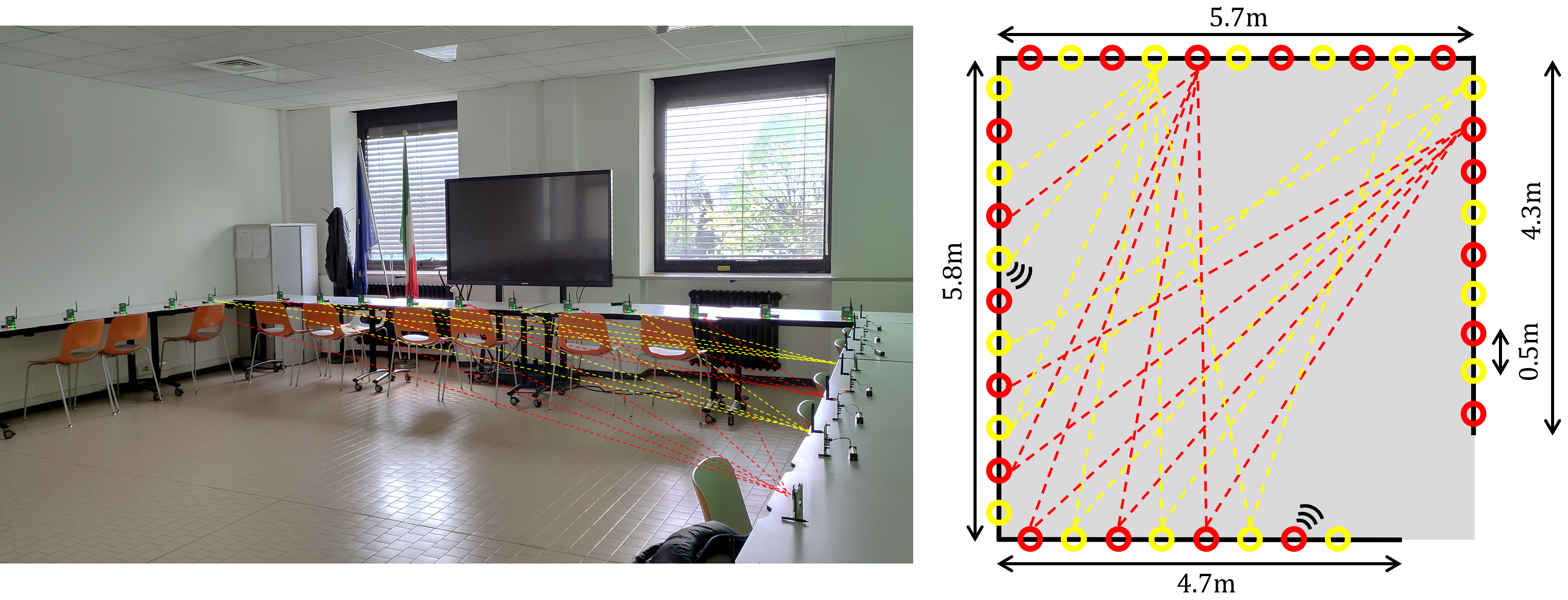} \caption{\label{tests} Case study scenario: left) monitored area; right) network layouts with main geometric parameters. Two networks (\#1 in red, and \#2 in yellow) are deployed in the same area. Note that the combined network (\#1+\#2) is not fully connected.}
\end{figure*}

The sensing area $\mathcal{X}$ is monitored by a network of radio nodes $\mathcal{V}$ connected through a set of active links $\mathcal{L}$. Each link $(u,v)=\ell \in \mathcal{L}$ defines a Fresnel's zone $\mathcal{F}_{u,v}$ (or, equivalently, $\mathcal{F}_{\ell}$), within which the presence of a human target significantly alters the received signal. We assume $N$ targets in the environment, each characterized by the physical parameter vector $\theta_n$ and located at position $\mathbf{P}_n = (x_n, y_n)$. A target is said to influence a link if at least 50\% of its elliptical body footprint lies within the Fresnel's zone of the link. The binary indicator function $\mathbb{1}_{(\cdot)}$ is used to define the presence of the $n$-th target in the Fresnel's zone $\mathcal{F}_{u,v}$ of the $\ell$-th link as:
\begin{equation}
\mathbb{1}_{\mathbf{P}_n \in \mathcal{F}_\ell} = 
\begin{cases}
1, & \text{if } \mathbf{P}_n \in \mathcal{F}_\ell. \\
0, & \text{otherwise}.
\end{cases}
\end{equation}
This equation, evaluated over all links, as shown in (\ref{eq:c2}), is used to form the set of all covering links for the $n$-th target:
\begin{equation}
Q_n := \{\ell \in \mathcal{L} \;|\; \mathbf{P}_n \in \mathcal{F}_\ell\}.
\end{equation}

Two necessary conditions are introduced to assess whether a target is resolvable:

\begin{itemize}
    \item \textbf{Link uniqueness condition (C1):} Targets must induce sufficiently distinct patterns of link occlusions. Given two targets $T_n$ and $T_m$, their respective link sets $Q_n$ and $Q_m$ must differ significantly to ensure distinguishability. This is quantified using the Jaccard distance \cite{Jaccard}:
\begin{equation}
\delta_{n,m} = 1 - \frac{|Q_n \cap Q_m|}{|Q_n \cup Q_m|}.\label{eq:jaccard}
\end{equation}
Thus, condition \textbf{C1} requires that, for all target indexes $m \neq n$,
\begin{equation}
\prod_{m \neq n} \mathbb{1}_{\delta_{n,m} > \tau} = 1,
\label{eq:c1}
\end{equation}
for a given tolerance threshold $\tau$. This ensures that $Q_n$ is not overly similar to any other target’s link set $Q_m$. The value of threshold $\tau$ depends on the geometry of the layout and the size of the targets: typical values lie in the interval $[0.2\,\,0.4]$.

\item \textbf{Minimum coverage condition (C2):} Each target must lie within the Fresnel's zone of at least one link to affect the network measurements. This is formalized as:
\begin{equation}
\sum_{\ell \in \mathcal{L}} \mathbb{1}_{\mathbf{P}_n \in \mathcal{F}_\ell} > 0 \quad \Leftrightarrow \quad |Q_n| > 0.
\label{eq:c2}
\end{equation}
\end{itemize}

Combining both conditions, the maximum number $\hat{N} \leq N$ of resolvable targets in the environment is given by:
\begin{equation}
\hat{N} = \sum_{n=1}^{N} \left( \prod_{m \neq n} \mathbb{1}_{\delta_{n,m} > \tau} \right) \cdot \mathbb{1}_{|Q_n| > 0} + \sum_{n=1}^{N} \frac{1}{\sum_{m \neq n} \mathbb{1}_{\delta_{n,m} \leq \tau}},
\label{eq:limit}
\end{equation}
where the first term counts the targets satisfying both conditions \textbf{C1} and \textbf{C2}, while the second term is a correction factor. This correction terms accounts for overlapping targets that cannot be individually distinguished due to identical or highly similar Fresnel's zone coverage. In such cases, indistinguishable targets contribute fractionally to the overall count, reflecting the limited spatial resolution of the sensing system.

This formulation provides a theoretical upper bound that depends explicitly on network parameters: number of links $L$, geometry of the node spatial deployment, and the RF carrier frequency (through Fresnel's zone size). Notably, the bound is valid under dense connectivity conditions, where $L \simeq |\mathcal{V}|(|\mathcal{V}| - 1)$.

The resolvability bound (\ref{eq:limit}) enables a model-based estimation of system capabilities, without requiring training data or ML tools. In Sect.~\ref{sec:Numerical-results-and}, we will use this framework in Monte Carlo simulations to evaluate how parameters like $\tau$, target density, and signal frequency influence performance. The bound also serves as a benchmark to compare with data-driven people counting strategies~\cite{depatla,kianoush-2016}.

\section{Experimental Validation and Results\label{sec:Numerical-results-and}}

This section evaluates the performance of the proposed RF sensing framework using both simulations and experimental data. The analysis includes theoretical bounds, synthetic RF attenuation samples generated via the diffraction models MAM and C-MAM, and a real-world indoor deployment in the 2.4\,GHz band.

\subsection{Simulation-Based Analysis}

\begin{figure*}[h]
  \centering
  \includegraphics[width=\textwidth]{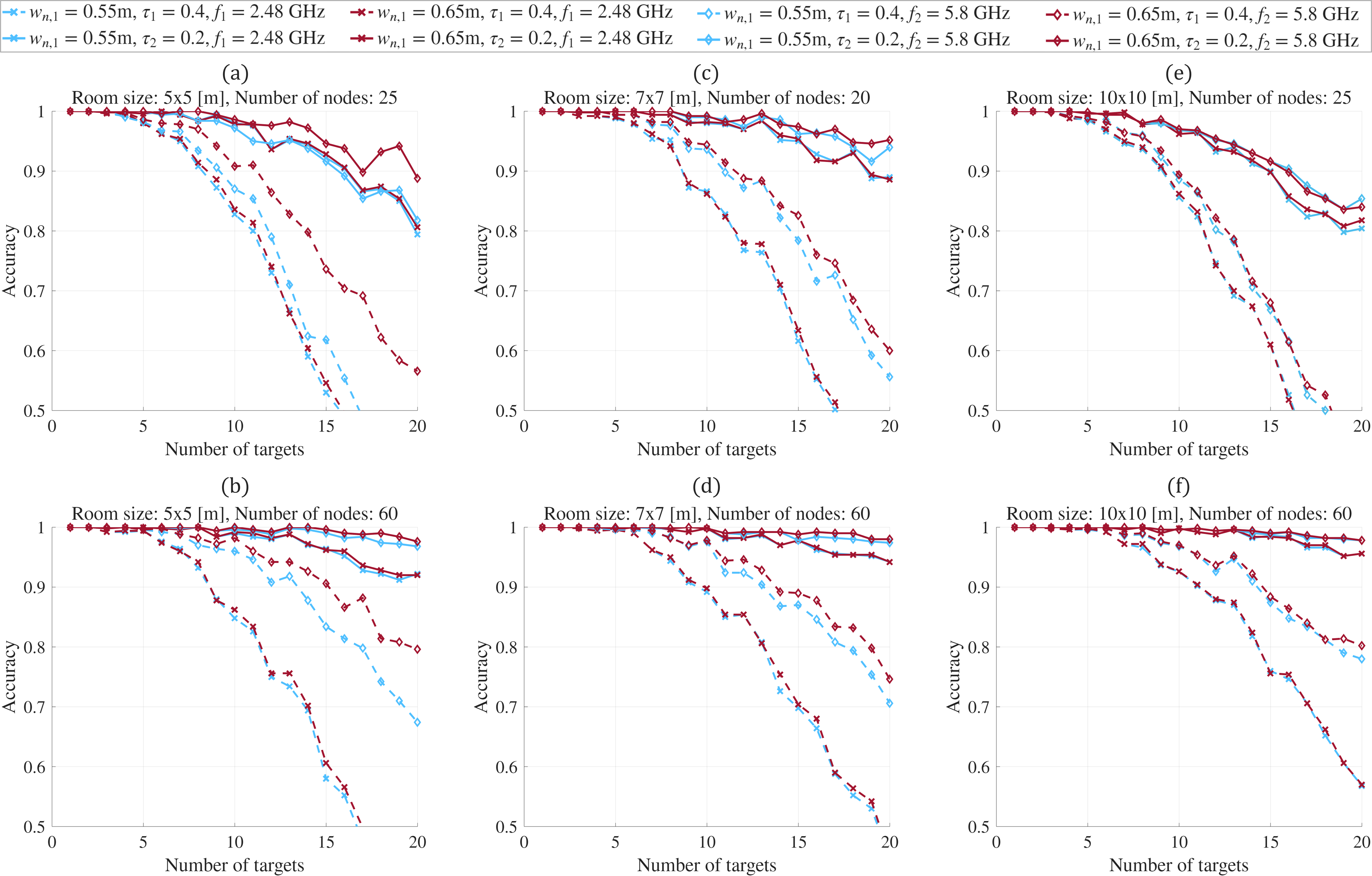}
  \caption{\label{fig4} Accuracy of the number of targets estimates as a function of the actual count, for different configurations and room sizes (5×5 m, 7×7 m, 10×10 m). Results derive from the theoretical bounds in Section~\ref{sec:limits}, considering: tuned Jaccard thresholds $\tau_1 = 0.4$ (dashed) and $\tau_2 = 0.2$ (solid), frequencies $f_1 = 2.48$ GHz (cross) and $f_2 = 5.8$ GHz (diamond), target widths $w_{n,1} = 0.55$ m (blue) and $0.65$ m (red), and number of nodes $|\mathcal{V}|=25/60$ (5×5 m), $|\mathcal{V}|=20/60$ (7×7 m), and $|\mathcal{V}|=60$ (10×10 m).}
\end{figure*}

We analyze the theoretical accuracy bounds (\ref{eq:limit}), computed as 
$
\frac{1}{n_{\text{s}}} \sum_{i=1}^{n_{\text{s}}} \mathbb{1}_{\hat{N}_i = N_i}.
$ Analysis is carried out under varying configurations corresponding to different room sizes (25~m$^2$, 49~m$^2$, 100~m$^2$), antenna densities (20, 25, 60 nodes), subject width $w_{n,1}$ (0.55~m, 0.65~m) and $w_{n,2}$ (0.25~m), subject height $h_n$ (1.6~m), operating frequencies (2.4~GHz, 5.8~GHz), and tuned Jaccard distance thresholds $\tau$ (0.2, 0.4). 

To generate the results shown in Fig.~\ref{fig4}, we performed Monte Carlo simulations based on the theoretical framework of Sect.~\ref{sec:limits}. For each configuration, we simulated up to \( N = 20 \) targets and, for every value of \( N \), we generated $n_{\text{s}}=500$ independent targets randomly placed in the monitored area. Then, we evaluated the resolvability conditions (\textbf{C1}) and (\textbf{C2}) to compute the estimated count \( \hat{N} \). Each point in Fig.~\ref{fig4} thus represents the average accuracy over $n_{\text{s}}$ trials, while the underlying network topology (node count and deployment geometry) and Fresnel's zone parameters (frequency, Jaccard threshold \( \tau \)) were fixed within each simulation batch.
As shown in Fig.~\ref{fig4}, the following trends are observed:

\begin{itemize}
    \item Higher antenna density significantly improves accuracy, particularly in large rooms.
    \item Larger subject width enhances target discriminability, especially when targets are sparsely distributed.
    \item Lower Jaccard thresholds ($\tau=0.2$) allow more targets to be resolved by enforcing stricter spatial separability.
    \item Operations at 5.8\,GHz yield better performance than at 2.4\,GHz, especially as $N$ increases.
\end{itemize}

With 60 antennas at 5.8 GHz, environments of 49--100 m$^2$ support accurate resolution of up to $N=14$--15 targets. Fewer nodes or larger inter-node spacing reduces the number of distinguishable targets to $N=6$--$10$, depending on the scenario.

\subsection{Experimental Validation}

\begin{figure}[h]
  \centering
  \includegraphics[scale=0.52]{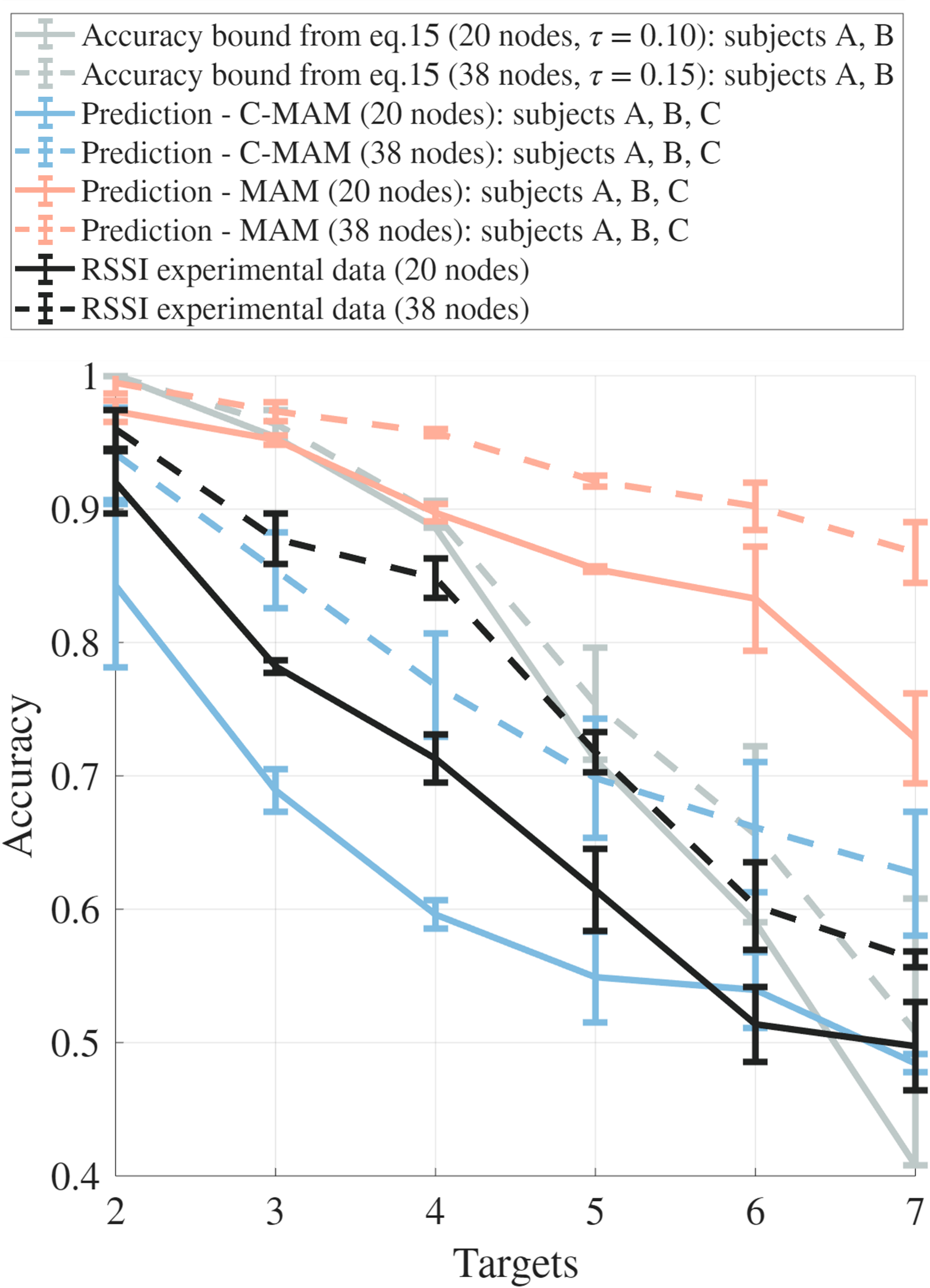}
  \caption{\label{last} Counting accuracy with RSSI data (black lines with error bars) for 20 (solid) and 38 (dotted) nodes, compared to predictions from C-MAM (blue) and MAM (pink) models. For C-MAM and MAM, error bars reflect subject variability (A, B, C). Accuracy bounds (\ref{eq:limit}) are shown in grey for both node settings, with Jaccard thresholds $\tau$ shown in the legend.}
\end{figure}

We conducted an experimental campaign using two IEEE 802.15.4-based battery-powered RF networks for IoT applications operating at 2.4\,GHz. Both networks were deployed over a 49 m$^2$ indoor area (Fig.~\ref{tests}), in accordance with typical industrial scenarios~\cite{davoli}. The two networks, represented as red and yellow nodes in Fig.~\ref{tests}, collect Received Signal Strength Indicator (RSSI) measurements simultaneously on different channels at 2.41 GHz and 2.43 GHz (IEEE channels 12 and 16), allowing approximation as a single network deployment due to negligible attenuation differences for very close frequencies~\cite{rampa-2017} (i.e., 20 MHz in our case).
 
Each network comprises 19 nodes, leading to a total of $|\mathcal{V}| = 38$ antennas with average spacing $\triangle = 0.5$ m and $L = 342$ active links. The received signal strength $P_{u,v}$ is measured for all links and used to compute the attenuation due to body presence according to:

\begin{equation}
P_{u,v}=\left\{ \begin{array}{ll}
P_{0}=P_{R}+w_{R} & \;\textrm{if \ensuremath{N=0}}\\
P_{1}=P_{R}-A_{u,v}+w_{T} & \;\textrm{if \ensuremath{N>0}, with subjects \ensuremath{\mathbf{\boldsymbol{\theta}}_{N}}},
\end{array}\right.
\label{eq:received_power}
\end{equation}

$P_0$ is the reference power (with no targets in the monitored area), and $w_R$, $w_T$ account for log-normal fading and noise~\cite{rampa-2022a,j-fieramosca-2023}. The body-induced attenuation is computed as $\hat{A}_{u,v} = P_{u,v} - \mathrm{E}_{t}[P_0]$, where $\mathrm{E}_{t}[\cdot]$ denotes the expected value over time. These per-link attenuation values are aggregated into vectors $\hat{\mathbf{A}}_{u,t}$ and collected across all nodes. These attenuation vectors are used as input features for a Regression Neural Network (RNN) trained on simulated data generated with the C-MAM model. The RNN outputs an estimate $\hat{N}$ of the number of subjects present. The network description is not detailed here due to the limited space available.

To investigate the effect of body size on the estimated results, we exploited three different subjects (Subject A, Subject B, and Subject C) having the following dimensions: 

\begin{itemize}
\item Subject A: $h_{n}=2$~m, $w_{n,1}=0.65$~m, $w_{n,2}=0.25$~m, 
\item Subject B: $h_{n}=1.6$~m, $w_{n,1}=0.55$~m, $w_{n,2}=0.25$~m, 
\item Subject C: $h_{n}=1.4$~m, $w_{n,1}=0.55$~m, $w_{n,2}=0.25$~m. 
\end{itemize}

We evaluate crowd counting accuracy using measured data with up to $N=7$ real subjects. Fig.~\ref{last} compares the empirical accuracy (black lines) to the predicted performance obtained using MAM and C-MAM models (pink and blue lines), both calibrated on the specific setup of Fig.~\ref{tests}. Accuracy bounds from (\ref{eq:limit}) are also included (grey lines) as a reference.

To evaluate the results across different body characteristics, MAM and C-MAM simulations were repeated for each of the three subject types (Subject A, Subject B, and Subject C), varying the number $N$ of simulated targets from 1 up to 7. Therefore, the accuracy values shown in Fig.~\ref{last} represent the average across these three subject types, while the error bars capture the attenuation variability introduced by the differences in body dimensions.

The key findings suggested by comparing the results of Fig.~\ref{last} are the following:

\begin{itemize}
    \item C-MAM predictions closely follow experimental accuracy, while MAM consistently overestimates performance for all values of $N$.
    \item For networks with 20 nodes, C-MAM slightly underestimates the accuracy when $N < 4$. This behavior is probably due to unmodeled multipath effects.
    \item For the 38-node setup, C-MAM aligns more accurately with the empirical data, though it tends to slightly overestimate accuracy for $N > 5$.
    \item Differences in body dimensions across subjects (Subject A, Subject B, and Subject C) contribute to higher accuracy variance as crowd size increases.
    \item The C-MAM model consistently predicts more conservative (lower) accuracy than MAM, avoiding non-physical attenuation summation in overlapping Fresnel's zones.
\end{itemize}

These findings confirm the practical relevance of the C-MAM model as a conservative yet effective tool for system performance estimation and deployment planning in RF sensing applications.

\section{Conclusion\label{sec:Conclusions}}

This paper presented an EM-informed approach for estimating the number of people in a monitored area using RF sensing within dense IoT network topologies. We introduced a Composite Multi-body Additive Model (C-MAM) that accounts for realistic electromagnetic interactions between RF signals and multiple human bodies, improving the physical fidelity of the attenuation modeling.
Based on this model, we derived practical resolvability limits that link the number of distinguishable targets to the geometric and electromagnetic properties of the network, including Fresnel's zone coverage and subject separability. These bounds enable a quantitative evaluation of system performance—such as average accuracy and maximum number of resolvable individuals—during the pre-deployment phase.
Our analysis considered a variety of propagation scenarios, room sizes, subject densities, and physical body dimensions. Theoretical predictions obtained with the C-MAM model were validated through experimental measurements of received signal strength in a controlled indoor setup, confirming the model’s relevance and predictive capability.
Future work will explore extending this framework to sparse and heterogeneous network configurations, and improving the estimation robustness for moving subjects, by keeping track of their movements using, for instance, Bayesian methods, and for changes in dynamic environments as well.

\end{document}